\documentclass[preprint,aps]{revtex4}

\usepackage{graphicx}
\usepackage{dcolumn}
\usepackage{bm}

\begin{document}


\title{ On the gauge invariant and topological nature of the localization\\
 determining the Quantum Hall Effect plateaus }
\author{Alejandro Cabo Montes de Oca }
\affiliation{Instituto de Cibernetica, Matematica y Fisica, Calle E
 309, Vedado, Ciudad Habana, Cuba}
\author{Danny Martinez-Pedrera}
\affiliation{Instituto de Cibernetica, Matematica y Fisica, Calle E
 309, Vedado, Ciudad Habana, Cuba}

\begin{abstract}
\bigskip
It is shown how the electromagnetic response of 2DEG under Quantum Hall
Effect regime, characterized by the
 Chern-Simons topological action,
transforms the sample impurities
 and defects in charge-reservoirs that
stabilize the Hall
 conductivity plateaus. The results, determine the basic
dynamical
 origin of the singular properties of localization under the
occurrence of the Quantum Hall Effect obtained in the pioneering
works of Laughlin and of Joynt and Prange, by means of a gauge
invariance argument  and a purely electronic analysis,
respectively. The common intuitive picture of electrons moving
along the equipotential lines gets an analytical realization
through the Chern-Simons current and charge densities.
\end{abstract}


\maketitle

Since its discovery in 1980, the integer Quantum Hall Effect (QHE)
has been the subject of a vast active research
\cite{klitzing,prange}. It is widely accepted that the explanation
of this rather remarkable effect, particularly the existence of
Hall conductivity plateaus, is intimately connected with
localization of electrons by sample impurities
\cite{prange,joynt,baraff}.

In the case of a perfect non-interacting two-dimensional electron
gas (2DEG) at sufficiently low temperature and with an exactly
integral number of Landau levels filled, the Hall conductivity
$\sigma_{yx}=n_eec/B=ne^2/h$, where $n$ is the integral number of
filled Landau levels, and this is a consequence of the well known
fact that each Landau level has a degeneracy $eB/hc$ per unit area
\cite{janssen}. For the Hall conductivity to show these quantized
values for wide ranges of the quantity $n_e/B$, which is indeed
the essential feature of the QHE, there must be some sort of
charge-reservoir in order to adjust the number of electrons in
extended current-carrying states to that required by a quantized
conductivity \cite{toyoda}. The charge-reservoir, it has been
argued, is provided by sample impurities or imperfections, and the
mechanism that of localization of electrons by the associated
random potential. It has been shown that as long as the Fermi
energy varies within a region of localized states the electrons in
extended states carry the right current for the Hall conductivity
to be quantized \cite{joynt}.

However, it is also recognized that a complete microscopic theory from
which the properties of
 the effect could be deductively obtained, is still
missing
 \cite{janssen,yoshioka}. The present work is intended to show
how the
electromagnetic response of the system, predicted by the field
theory considered in \cite{caboliva,randjbar}, can sustain a
microscopic description from which some basic properties of the role of
impurities for QHE
 could be derived.
Here,  it is evidenced that the electromagnetic response of the
system, described by the Chern-Simons topological action,
effectively transforms the sample impurities in charge-reservoirs.
Moreover, for some ranges of variation of applied magnetic field, the system
is  capable of adjusting the electron density in most of the sample
points to those values required for satisfying the integral filling of
Landau levels locally. This picture gives, thus,  a fundamental explanation
of the results of references  \cite{laughlin,joynt}, highlighting the
gauge invariance and topological character of the effect. The results also
lead ground to  the applicability of the field description of integer QHE
presented in \cite{cabochaichian}, where the mechanism of chemical potential
stabilization in a gap remained unclear.

To simplify the discussion the analysis has been carried on a
multi quantum well structure (superlattice) \cite{haavasoja,stormer}, where
the field distribution of interest  can be obtained analytically. 
First, the field distribution associated to a model impurity in
the superlattice is presented and its connection to the
Chern-Simons topological action is shown. Next, it is discussed
the stability of the QHE and the emergence of the Hall
conductivity plateaus is seen.

The electromagnetic response $a_{\mu}(x)$ represents a linear
disturbance of the electromagnetic potential associated to the
presence of the constant magnetic field $B_n=n_ehc/ne$ at which
the filling factor $\nu$ has precisely the value $n$. In
\cite{caboliva} it was found the equation satisfied by the
electromagnetic response of a superlattice in QHE regime
\begin{eqnarray}
\partial^2a_{\mu}(x)+i{\frac{4\pi\sigma_H}{ca}}\epsilon^
{\alpha\mu\sigma\nu}n_{\alpha}\partial_{\sigma}a_{\nu}(x)\nonumber\\
+4\pi{\frac{\chi_e}a}\bigl[P_{\mu\nu}u_{\alpha}u_{\beta}+
u_{\mu}u_{\nu}P_{\alpha\beta}\nonumber\\
-(u_{\mu}P_{\nu\alpha}+u_{\nu}P_{\mu\alpha})u_{\beta}\bigl]
\partial_{\alpha}\partial_{\beta}a_{\nu}(x)=0,
\label{eq:eq}
\end{eqnarray}
which, in order to evidence the topological Chern-Simons terms in
it, has been written in Lorentz-covariant form. Also, $a$ is
the distance between the 2DEG planes in the superlattice,
$u_{\mu}=(1,0,0,0)$ is the superlattice 4-velocity,
$n_{\mu}=(0,0,0,1)$ is a unitary 4-vector normal to the 2DEG
planes, $P_{\mu\nu}=\mathrm{diag}(0,1,1,0)$ is the projection
tensor on the 2DEG planes, and $\sigma_H=ne^2/h$ and
$\chi_e=n{\frac{e^2}h}{\frac{mc}{eB_n}}$ are the Hall conductivity
and the dielectric susceptibility of a single plane of electrons
at filling factor $\nu=n$, respectively.
The 4-potential $a_{\mu}(x)$
 is taken in Coulomb gauge and will be
supposed independent of $x_3$ coordinate. It will also be assumed
that $a_3=0$.

The effective Maxwell equations (\ref{eq:eq}) were obtained in
\cite{caboliva} within the long wavelength approximation
$\lambda\gg r_0=\sqrt{\frac{\hbar c}{eB_n}}$ from a calculation of
the first quantum correction to the effective action of a 2DEG in
the presence of a magnetic field. The equations appropriate to a
layered multi quantum well structure are obtained by simply adding
the equations for each plane. The above equations (\ref{eq:eq}) 
also can be alternatively derived  by extremizing with
respect to $a_{\mu}(x)$ the effective action
\begin{eqnarray}
\Gamma_\mathrm{eff}[a_{\mu}]=\int d^2x\,dt\left[{\frac 1{8\pi}}
\epsilon E^2-{\frac 1{8\pi}}B^2\right.
\left.+{\frac{i\sigma_H}{4ca}}\epsilon^{\alpha\mu \nu}a_{\alpha}
F_{\mu\nu}\right],
\label{eq:effac}
\end{eqnarray}
where $\epsilon=1+4\pi{\frac{\chi_e}a}$,
$F_{\mu\nu}=\partial_{\mu} a_{\nu}-\partial_{\nu}a_{\mu}$,
$E_j=-iF_{0j}$ and $B={\frac 12} \epsilon^{jk}F_{jk}$ $(j,k=1,2)$.
The third term in the integrand of (\ref{eq:effac}) is recognized
as the Chern-Simons topological action, since is independent of
the metric tensor.

In the present  work, the impurity in superlattice sample will be
modelled as a
 cylindrical hole of radius $\eta$ having the  axis normal to
the 2DEG.
 Then, we search for stationary axially-symmetric solutions to
(\ref{eq:eq}),
 which in this case reduces to consider the following equations
for the
 scalar potential $\phi(r)$ and the azimuthal component of the
vector potential $a_{\theta}(r)$
\[{\frac d{dr}}\left(r{\frac{d\phi}{dr}}\right)=-{\frac{4\pi\sigma _H}{ca}}
\frac{1}{\epsilon}{\frac d{dr}}(ra_{\theta}),\]
\[{\frac 1r}{\frac d{dr}}\left(r{\frac{da_{\theta}}{dr}}\right)-
{\frac{a_{\theta}}{r^2}}={\frac{4\pi\sigma_H}{ca}}{\frac{d\phi }{dr}}.\]
These equations were investigated in \cite{guevara}. A solution,
finite for $r\to \infty,$ is given by
\begin{eqnarray}
\phi (r)={\frac C{\sqrt{\epsilon}}}K_0(kr),\nonumber\\a_{\theta}(r)=
C\left(-{\frac 1{kr}}+K_1(kr)\right),
\label{eq:solpot}
\end{eqnarray}
where $C$ is an overall constant factor to be determined, $K_0(x)$
and $K_1(x)$ are MacDonald functions of $0\mathrm{th}$ and
$1\mathrm{st}$ order respectively, and $k={\frac{4\pi \sigma
_H}{ca\sqrt{\epsilon}}}$. From these expressions, the response
electromagnetic field of the superlattice is readily obtained as
\begin{eqnarray}
E(r)={\frac{kC}{\sqrt{\epsilon }}}K_1(kr), B(r)=-kCK_0(kr).
\label{eq:solfld}
\end{eqnarray}
Note that the magnetic field is proportional to the electric
potential. These fields decay exponentially for $r\to \infty $,
being $k^{-1}$ a measure of its effective extension, and will be
adopted as valid for the region external to the cylinder
 $r\geq
 \eta.$ Within the interior, for completely defining the impurity
model, the electric potential will be assumed constant and the
magnetic intensity will also be assumed to be a constant vector
along the axis orthogonal to the superlattice planes.

Using the expression (\ref{eq:solfld}) for $B(r)$, the magnetic
flux $\Phi $ associated to the impurity can be calculated to be
\begin{eqnarray}
\Phi =\int \!\vec B \cdot \vec n dS=\int_{\eta}^{\infty} \!B(r)2\pi rdr
=-2\pi C\eta K_1(k\eta ),
\end{eqnarray}
which in the limit $\eta \rightarrow 0$
\begin{equation}
\lim_{\eta \rightarrow 0}\Phi =-2\pi {\frac Ck}.
\label{eq:Phi}
\end{equation}
Then, the edge current which  flows through the boundary of
the cylinder should be such that in conjunction with the
continuous distributed Hall currents will produce the just
determined total magnetic flux. The explicit determination of this
current is not necessary for the further steps of constructing the
model.

Let us now consider the edge electric charge. Applying Gauss law
to the cylinder, the free edge charge $q$ per unit length in $x_3$
direction in the impurity can be obtained as

\begin{equation}
q={\frac 12}\eta \epsilon E(\eta ^{+})=\frac 12 C\sqrt{\epsilon}
k\eta K_1(k\eta ),
\end{equation}
and

\begin{equation}
\lim_{\eta\to 0} q=\frac 12 C\sqrt{\epsilon}.
\label{eq:q}
\end{equation}
After substituting the solution (\ref{eq:solpot}) in the
expression (\ref{eq:effac}) for the effective action we get
\begin{equation}
\Gamma_\mathrm{eff}={\frac{C^2}4}K_0(k\eta )\int \!dt.
\label{eq:evaleffac}
\end{equation}
Using the fact that, for stationary equilibrium fields, the
effective action is minus the space-time integral of the energy
density, considering the contribution to the energy of the free
edge charge, and using (\ref{eq:solpot}), (\ref{eq:q}) and
(\ref{eq:evaleffac}), we obtain, for the mean energy density
$\mathcal{U}$ associated to an impurity, the expression
\begin{equation}
\mathcal{U}=-{\frac{\Gamma_\mathrm{eff}}{A\int \!dt}}+
{\frac{q\phi (\eta^{+})}{A}}={\frac{C^2}{4A}}K_0(k\eta ),
\label{eq:U}
\end{equation}
where $A$ is the sectional area of the sample. This is the energy
required to produce a static deformation (\ref{eq:solpot}) of the
integral filling homogeneous field $B_n$.

As the next step in the definition of the model, we will consider
$N$ impurities like that described above, spread over the sample
area to a mean separation $\xi \equiv \sqrt{A/N}$. The  holes will
also be assumed to be small for having a total area very much
smaller than the sample surface. Hence, the electromagnetic mean
field associated to the inspected many electron state, should be
approximately given by  superposing  the fields of all the
impurities. An illustrative picture of the spacial dependence on
the 2DEG planes of the electric potential is shown in Fig.1.  At
this point, it should be noticed that the field solving the
Maxwell equations (1), exactly satisfy the following properties:
a) The Hall current always flows along the equipotential level
curves. b) The normal magnetic field to the planes and the charge
densities  at any point out of the cores of the impurities also
exactly satisfies the integral filling condition, as it was
discussed in \cite{guevara}. Therefore, it can be concluded, that
the electric potential and the Hall currents of the considered
many electron state, analytically  realize the usual intuitive
picture for the conduction in QHE and moreover, also gives ground
for early proposed percolation models \cite{prange,yoshioka}.

\begin{figure}[ht]
\includegraphics[width=3in]{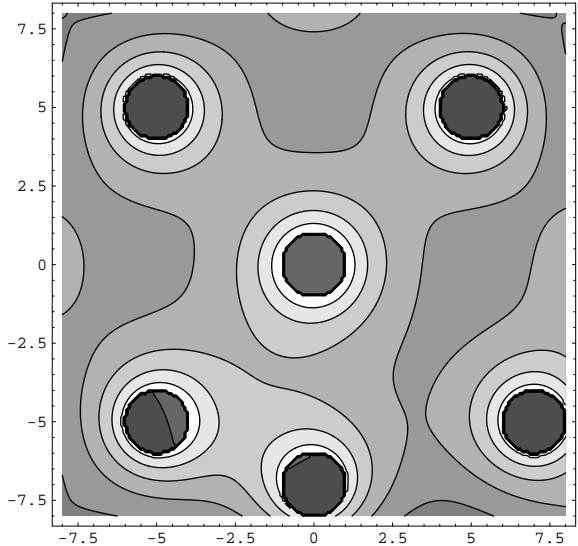}
\caption{\label{fig:plateaus} Contour plot illustrating the
equipotential  curves of the superposed fields for various
impurities. The Hall currents flow along them, and  while some
lines are closed, other ones connect different spatial regions
reflecting the presence of extended states.}  \end{figure}

Let us resume the picture we intend to support in the argue to be
done below. Whenever the external magnetic field $B_\mathrm{ext}$
slightly deviates from one of the values required by integral
filling of Landau levels, each impurity accumulates a free edge
charge $q$ such that the electron density at any internal point in
the sample volume, adjusts to that required to satisfy the
integral filling condition at the precise local magnetic field
value.  For the sample type we have been considering, the response
electromagnetic field is exponentially damped away from the
impurities.  Therefore, the excess of magnetic flux over that
corresponding to integral filling flux density $B_n$ will  tend to
concentrate more around the impurities. If the distance between
them turns to be greater than the penetration length, then as
mentioned before, a sort of Meissner effect occurs. Then,the
difference between the external flux and the one associated to a
complete filling of some number of Landau levels will be expelled
from the large volumes and concentrated around the defects
\cite{caboliva}. The defect properties will resemble in such
regimes the vortices in type II superconductors. This should not
be necessarily the case.  It also should be stressed that,
although in the realistic samples, the defects can be expected  to
be distributed in uncorrelated positions on each plane of the
superlattice, it is natural to suppose  that the magnetic flux
lines of the real defects will follow trajectories that traverse
the sample in $x_3$ direction with some  distortion respect (like
the real Abrikosov vortices in non-ideal samples) to the perfect
line configuration we are considering.

In order to justify the above picture,  it should be shown that the
system energy is lower than the value corresponding to the purely
homogeneous field configuration. Let us consider this problem in
what follows.

To start, let us define the connections between the main parameters
characterizing the model. From the condition of equality of total
magnetic flux in and out of the sample,
\[B_\mathrm{ext}A=B_nA+N\Phi,\]
and from the expression (\ref{eq:Phi}), the constant $C$ is determined to be
\[C=-\frac{B_\mathrm{ext}-B_n}{2\pi}k\xi^2.\]

Substituting this result in expression (\ref{eq:U}) and
considering there are $N$ impurities, it is
obtained the energy density in QHE regime
\begin{equation}
\mathcal{U}_\mathrm{QHE}(B_\mathrm{ext})=\frac{(B_\mathrm{ext}-B_n)^2}
{16\pi^2}k^2\xi^2K_0(k\eta)
\label{eq:Uqhe}
\end{equation}

It must be remarked, that here, in order to simplify the
discussion, for evaluating this energy, it was assumed that the
field distributions of different impurities do not overlap very
much. This supposition was employed in order to approximately
calculate the energy as the simple sum of the contributions
associated to  each defect. A more precise procedure is however,
possible for the search of more quantitative results.

On another hand, when the homogeneous field $B_\mathrm{ext}$ completely
penetrates the sample, the total energy density required for such deformation
over the integral filling field $B_n$ is calculated to be
\begin{equation}
\mathcal{U}_\mathrm{homog}(B_\mathrm{ext})=\frac{(B_\mathrm{ext}-B_n)^2}
{8\pi}+\mathcal{U}_\mathrm{Peierls}(B_\mathrm{ext}),
\label{eq:Uhomog}
\end{equation}
where $\mathcal{U}_\mathrm{Peierls}$ is a contribution to the
energy due to Peierls and described in \cite{randjbar}; it is
given by
\begin{widetext}
\begin{equation}
\mathcal{U}_\mathrm{Peierls}(B_\mathrm{ext})=\frac{{n_e}^2h^2}{4\pi
ma}\left\{-{\left[\frac{B_1}{B_\mathrm{ext}}\right]}\left({\left[
\frac{B_1}{B_\mathrm{ext}}\right]}+1\right){\left(\frac{B_\mathrm{ext}} 
{B_1}\right)}^2+\left(2{\left[\frac{B_1}{B_\mathrm{ext}}\right]}+1\right)
\frac{B_\mathrm{ext}}{B_1}-1\right\},
\end{equation}
\end{widetext}
where $[x]$ denotes the integer part of $x$.


For the QHE regime to be energetically favorable it is clear we must have
\[
\mathcal{U}_\mathrm{QHE}(B_\mathrm{ext})<\mathcal{U}_\mathrm{homog}(B_\mathrm{ext})
\]
and this is equivalent to
\begin{eqnarray}
\left(\frac{k^2\xi^2K_0(k\eta)}{2\pi}-1\right)\frac{(B_\mathrm{ext}-B_n)^2}{8\pi}
 <\mathcal{U}_\mathrm{Peierls}(B_\mathrm{ext}).
\label{eq:qhecond}
\end{eqnarray}
As is easily seen, the QHE regime is always preferred whenever
$\xi<\frac 1k\sqrt{2\pi/K_0(k\eta)}$, which means that the Hall
conductivity shows a perfect step-like dependence. However, in the
limit of a very small $\xi$ the impurities will strongly interact
and the picture would no longer be applicable. In any case, this
dirty limit should eventually destroy the QHE regime. When
$\xi>\frac 1k\sqrt{2\pi/K_0(k\eta)}$ the inequality
(\ref{eq:qhecond}) predicts a plateau width
\[
(\Delta
B)_n=B_n\left\{\frac{2n(n^2+\alpha)}{(n^2+\alpha)^2-n^2}\right\}
\]
where we have denoted
$\alpha=\frac{mac^2}{2e^2}\left(\frac{k^2\xi^2K_0(k\eta)}{2\pi}-1\right)$.
This result would clearly express the QHE stability if there are
no alternative states of the system showing less energy.

The plateaus predicted by the above expressions, have been shown
in Fig.~\ref{fig:plateaus} for four values of the quantity
$\alpha$. We have used the same values for the physical parameters
as those of the experiments described in \cite{haavasoja}.

\begin{figure}[ht]
\includegraphics[width=3in]{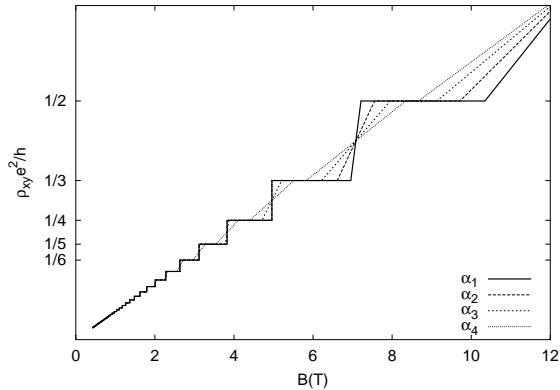}
\caption{\label{fig:plateaus} Predicted plateaus for four values
of the quantity $\alpha$.}
\end{figure}

Summarizing, it has been illustrated that the electromagnetic
response of a superlattice of 2DEG free of impurities is able to
transform an added distribution of impurities or defects in a set
of charge reservoirs. The excess charge over that required by an
integral filling of Landau levels is then accumulated in the
impurities.   It can be understood  from
the present analysis, that it is the dynamically acquired capacity
of accumulating charge of the impurities the main element to
account for this effect. A next important step of the present
study would be to investigate a similar model but in the case of
planar samples. Analogous results can be expected in this case.
The main additional complication seems to be the determination of
the fields associated to a single defect, since that for that
situation these fields spray out in the 3D space. 
 An additional task for this
planar case, would be the inclusion of the temperature. 
We expect this study to provide a
foundation  of the successful  phenomenological model of Ingraham
and Wilkes \cite{ingraham}. In addition, the present discussion
can also be viewed  as giving a microscopic explanation of the
model of T. Toyoda et al \cite{toyoda}. These authors assumed the
existence of a kind of particle reservoir being in equilibrium
with the 2DEG and with a limited capacity to account for the
plateau widths found in experiments. The fact that the total
energy (\ref{eq:Uqhe}) becomes greater than the energy of the
homogeneous magnetic field distribution can be expected to play
the role of the limiting capacity stopping the particle
accumulation and accounting for the plateau widths.

We acknowledge,
the AS ICTP Associate and Federation Schemes for the helpful
support in the developing of this work.Our recognition is also expressed  to
R. Guevara for his
 participation in conceiving the idea of article and to
Profs.
 N.H. March and G. Thomson for their very valuable comments at
ASICTP .

\end{document}